\documentclass{emulateapj}
\usepackage{natbib}
\usepackage{apjfonts}

\usepackage{amsmath}

\usepackage{epsfig}
\usepackage{graphicx}
\usepackage{amssymb}
\usepackage{aas_macros}
\usepackage{color}

\newcommand{\oversim}[2]{\protect{\mbox{\lower0.5ex\vbox{%
   \baselineskip=0pt\lineskip=0.2ex
   \ialign{$\mathsurround=0pt #1\hfil##\hfil$\crcr#2\crcr\sim\crcr}}}}} 

\catcode`\"=\active\let"=\"  

\def\3{{\ss} }

\def\c12{{1\over 2}}

\def\d{{\rm d}}   
   
\def\plusplus{\raise 0.3ex\hbox{${\scriptstyle ++}$}{}}

\begin{document}   
   
\title{The Coupling between the core/cusp and missing satellite problems} 
\author{Jorge Pe\~{n}arrubia\altaffilmark{1,2}, Andrew Pontzen\altaffilmark{3}, Matthew G. Walker\altaffilmark{4} \& Sergey E. Koposov\altaffilmark{2,5}}

\email{jorpega@iaa.es}
\altaffiltext{1}{Ram\'on y Cajal Fellow, Instituto de Astrof\'isica de Andalucia-CSIC, Glorieta de la Astronom\'ia, 18008, Granada, Spain}
\altaffiltext{2}{Institute of Astronomy, University of Cambridge, Madingley Road, Cambridge, CB3 0HA, UK}
\altaffiltext{3}{James Martin Research Fellow, Oxford Astrophysics, Univ. of Oxford, Denys Wilkinson Bldg., Keble Road, OX1 3RH, UK}
\altaffiltext{4}{Hubble Fellow, Harvard-Smithsonian Center for Astrophysics, 60 Garden St., Cambridge, MA 02138}
\altaffiltext{5}{Sternberg Astronomical Institute, Moscow State University, Universitetskiy pr. 13, 119992 Moscow, Russia}


\begin{abstract} 
We calculate the energy that baryons must inject in cold dark matter (CDM) haloes in order to remove centrally-divergent DM cusps on scales relevant to observations of dwarf spheroidal galaxies (dSphs). We estimate that the CDM haloes often associated with the Milky Way's dSphs ($M_{\rm vir}/M_\odot\sim 10^{9-10}$) require $\Delta E\sim 10^{53-55}$erg in order to form cores on scales comparable to the luminous size of these galaxies. 
While supernova type II (SNeII) explosions can in principle generate this energy, the actual contribution is limited by the low star formation efficiency implied by the abundance of luminous satellites. Considering that CDM's well-known `core/cusp' and `missing satellite' problems place opposing demands on star formation efficiencies, existing observational evidences for large cores in the most luminous dSphs require that CDM models invoke some combination of the following: (i) efficient (of order unity) coupling of SNeII energy into dark matter particles, (ii) star formation histories peaking at unexpectedly high redshifts $(z\gtrsim 6$), (iii) a top-heavy stellar IMF, and/or (iv) substantial satellite disruption or other stochastic effects to ease the substructure abundance constraints. Our models show that the tension between CDM problems on small scales would increase if cored DM profiles were to be found in fainter dwarves. 
\end{abstract}   

\section{Introduction} \label{sec:intro}
Cosmological N-body simulations show that if gravitational
interactions between `standard' (i.e., massive, weakly-interacting) cold
dark matter (CDM) particles dominate structure formation, then galaxies must be embedded in dark matter haloes that (i) collectively follow a mass function that diverges at low masses as $\d N/\d M_{\rm vir}\sim M_{\rm vir}^{-1.9}$ (e.g. Springel et al. 2008 and references therein); and (ii) individually follow mass-density profiles characterized by centrally-divergent `cusps', with $\rho(r)\propto r^{-1}$ at small radii (Dubinski \& Carlberg 1991; Navarro, Frenk \& White 1996, hereafter NFW). 
The dark matter haloes inferred from observations of real galaxies differ significantly: the estimated mass profiles are consistent with homogeneous-density 'cores' (Kuzio de Naray et al. 2008; Battaglia et al. 2008; de Blok 2010; Walker \& Pe\~narrubia 2011; Amorisco \& Evans 2012) and the satellite mass function is remarkably flat (Klypin et al. 1999; Moore et al. 1999).  

Either the dark matter is not standard CDM or, if it is, then
non-gravitational forces must play a significant role in structure
formation.  Indeed various baryon-physical mechanisms have
been proposed and demonstrated to be capable of fixing both problems {\it separately}: 
e.g. supernova explosions (Navarro et al. 1996, Read \& Gilmore 2005; Mashchenko et al. 2008; Governato et al. 2010; Pontzen \& Governato 2012), or the orbital decay of compact baryonic objects (El-Zant et al. 2001; Goerdt et al. 2010; Cole et al. 2011) can under plausible conditions flatten the central cusps of CDM-like haloes.  Likewise, supernova feedback, inefficient cooling of interstellar media, cosmic reionization, UV radiation, and acoustic oscillations have all been invoked to explain the suppression of galaxy formation in low-mass satellite haloes
(e.g. Tassis et al. 2008; Bovill \& Riccotti 2009; Sawala et al. 2010; Bovy \& Dvorkin 2012).

Notice that solutions to CDM problems on small scales rest upon the
efficiency with which stars form in DM haloes. However, while
core formation generally requires an efficient conversion of gas into stars,
suppression of galaxy formation obviously requires the opposite.

In light of this apparent tension, here we examine whether baryon-physical
solutions to CDM's `core/cusp' and `missing satellites' problems are
mutually compatible. 
For maximal leverage we consider explicitly the
impact of the most energetically efficient baryon-driven
events--supernova explosions of type II (SNeII)\footnote{One may show that the orbital energy of compact stellar objects is orders of magnitude lower than associated to SNeII events.} --on the least
luminous objects associated empirically with dark matter haloes: the Milky
Way (MW)'s dwarf spheroidal satellites.


\section{Cusp removal at work}\label{sec:work}

\subsection{Models}\label{sec:transf}
We proceed to derive a rough estimate of the amount of work involved in transforming a centrally-divergent cusp into a constant-density core. We adopt the following cosmologically-motivated halo density profile
\begin{eqnarray}
\label{eq:rho}
\rho(r)=\frac{\rho_0 r_s^3}{(r_c+r)(r_s+r)^2};
\end{eqnarray}
where $\rho_0$ is a characteristic halo density and $r_s$ is a scale radius. We assume that the original state of the DM follows a NFW model\footnote{Adopting Einasto profiles instead of NFW models has a negligible impact on our estimates.}, i.e. for a core radius $r_c= 0$, and that the main effect of baryonic feedback is to flatten the inner cusp, so that at $r\ll r_c$ dark matter is homogeneously distributed. Clearly, this solution is not unique, and in principle more complex models (that is, with a larger number of free parameters) can be invoked in order to describe the effects of baryonic feedback in more elaborate a detail. In practice, however, eq.~(\ref{eq:rho}) provides a reasonable fit to the cored DM density profiles found in the hydro-dynamical simulations of Governato et al. (2012, G12). Also, one may show that the below estimates barely change for more complex profiles, e.g $\rho/\rho_0\sim (r^n +r_c^n)^{-1/n}$ at $r\ll r_s$.

After the DM cusp is removed we assume that the halo settles in a new equilibrium state. Hence, we can use the virial theorem to derive a {\it lower} (i.e conservative) limit of the amount of energy required by the transformation as $\Delta E=\Delta W/2= (W_{\rm core}-W_{\rm cusp})/2$, where $W$ is the halo potential energy 
\begin{eqnarray}
\label{eq:W}
W=-4 \pi G \int_0^{r_{\rm vir}} \rho(r)M(r)r\d r;
\end{eqnarray}
and $M(r)$ is the halo mass profile, which can be integrated analytically from eq.~(\ref{eq:rho})

\begin{equation}
\frac{M(r)}{M_0}= 
\begin{cases}
\frac{x^2 \ln(1+\tilde r/x)+(1-2x)\ln(1+\tilde r)}{(1-x)^2} -\frac{\tilde r}{(1+\tilde r)(1-x)} & ,x\ne 1 \\ 
\ln(1+\tilde r) - \frac{\tilde r(2+3\tilde r)}{2(1+\tilde r)^2} & , x=1 
\end{cases}
\label{eq:m}
\end{equation}
where $M_0\equiv 4 \pi \rho_0 r_s^3$, $\tilde r \equiv r/r_s$ and $x\equiv r_c/r_s$.

Given that feedback mainly shapes the inner-most regions of DM haloes, the parameter $\rho_0$ can be derived from eq.~(\ref{eq:m}) under the assumption that the halo virial mass remains constant\footnote{However, in general we expect haloes to grow in mass while forming stars, i.e $M_{\rm vir,core}\gtrsim M_{\rm vir, cusp}$, hence our estimates of $\Delta E$ must be regarded as conservative.}, i.e. $M_{\rm vir, core}=M_{\rm vir, cusp}=M_0[\ln(1+c)-c/(1+c)]$, where $c=r_{\rm vir}/r_c$ is the halo concentration.




It is useful to write
\begin{eqnarray}
\label{eq:delw}
\Delta W= \frac{G M_{\rm vir}^2}{r_s} \Theta(x,c).
\end{eqnarray}
Although this equation has to be solved numerically, one may show that $\Theta\sim x$ for $x\ll 1$, i.e. the energy required to form small cores is proportional to the core size; whereas for $x\gg 1$, this function approaches asymptotically the limit $\lim_{x\rightarrow \infty}\Theta\approx 0.05$. This indicates that unbinding the very central cusp is the energetically demanding part of the transformation, whilst re-distributing DM beyond the halo scale radius requires, in comparison, a relatively small amount of energy, suggesting that our results will not be strongly sensitive to the details of the outer halo profile.

In our models the initial state of the halo is fully specified by $M_{\rm vir}, r_{\rm vir}$ and $c=r_{\rm vir}/r_s$. These parameters can be fixed by using the results from N-body cosmological simulations. For example, the halo virial radius $r_{\rm vir}=r_{\rm vir}(M_{\rm vir},z)$ is calculated from Bullock et al. (2001), whereas the concentration $c=c(M_{\rm vir},z)$ follows the relation proposed by Macci{\`o} et al. (2007)\footnote{The scatter in this relation $\pm 0.15$ dex is neglected for reasons that will become obvious below.}. Hence the amount of work required to remove the DM cusp ($\Delta W$) is solely determined by the size of the DM core with respect to the original scale radius, $x=r_c/r_s$.


\begin{figure}
\includegraphics[keepaspectratio=true,height=84mm]{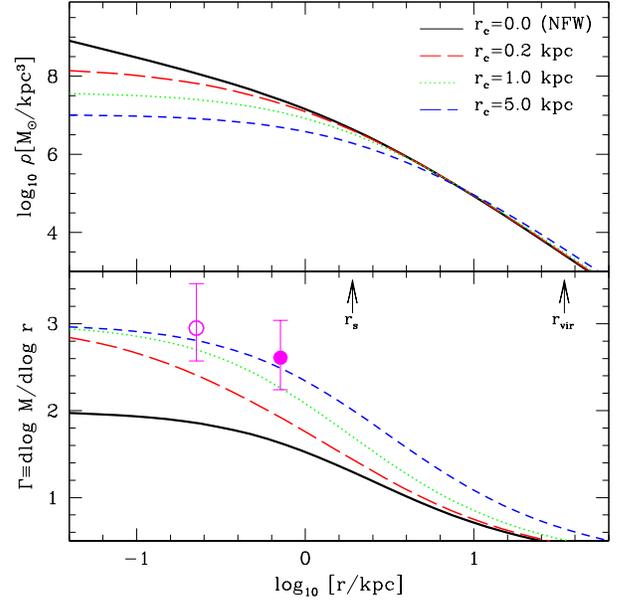}
\caption{{\bf Upper panel:} Density profiles considered in this work. Thick solid lines correspond to an NFW model with $M_{\rm vir}=3\times 10^9M_\odot$, scale radius $r_s\simeq 1.9$ kpc a concentration $c=16$. Red, green and blue show cored density profiles of haloes with the same virial mass and a core radius $r_c=0.2, 1.0$ and 5 kpc, respectively.  {\bf Lower panel:} Slope of the mass profile $\Gamma\equiv \d\ln M/\d\ln r$ as a function of radius for the above models. NFW and cored haloes have $\Gamma<2$ and $\Gamma<3$, respectively. Recent measurements of $\Gamma$ in the Fornax (closed symbol, $r_{\rm h}\simeq 713$ pc) and Sculptor (open symbol, $r_{\rm h}\simeq 226$ pc) dSphs are indicative relatively large ($r_c\gtrsim 1$~kpc) DM cores in both galaxies (see WP11 for details).}
\label{fig:prof_slope}
\end{figure}

\subsection{Core size}\label{sec:coresize}
Recently, Walker \& Pe\~narrubia (2011; WP11) have measured the slope of the halo mass profile ($\Gamma$) in the two brightest MW dSphs, Fornax and Sculptor ($M_\star/M_\odot\simeq 3\times 10^7$ and $8\times 10^6$ respectively; de Boer et al. 2012a,b). The values, $\Gamma\equiv \Delta \log M / \Delta \log r=2.61_{-0.37}^{+0.43}$ and $\Gamma=2.95_{-0.39}^{+0.51}$, rule out NFW profiles ($\Gamma< 2$) at confidence levels $\gtrsim 96\%$ and $\gtrsim 99\%$, respectively.

We can use eq.~(\ref{eq:rho}) to relate the slope of the mass profile measured at the half-light radius ($r_{\rm h}$) with the core size. For stellar components that are deeply embedded within the DM halo ($r_{\rm h}\ll r_s$) we have
\begin{eqnarray}
\label{eq:gamma}
\Gamma(r_{\rm h}) = 3 -\frac{3(1+2x)}{4x}\bigg(\frac{r_{\rm h}}{r_s}\bigg)+\mathcal{O}\bigg(\frac{r_{\rm h}}{r_s}\bigg)^2.
\end{eqnarray}
Two points can be gleaned from this Equation. First, measurements of $\Gamma(r_{\rm h})$ cannot be used to derive upper limits for the halo core radius, as the slope approaches an asymptotic value $\Gamma\simeq 3 - 3/2(r_{\rm h}/r_s)$ for $r_c\gg r_s$. And second, steep mass profiles ($\Gamma\gtrsim 2.5$) imply that the dark matter core extends well beyond the luminous radius of the dwarf, i.e $r_c\gtrsim r_{\rm h}$.

Fig.~\ref{fig:prof_slope} illustrates the transformation defined by eq.~(\ref{eq:rho}). We adopt a fiducial NFW model for dSphs with $M_{\rm vir}=3\times 10^9M_\odot$, $r_s\simeq 1.9$ kpc and $c=19.6$ (solid line) (see Pe\~narrubia et al. 2008). Long-dashed, dotted and short-dashed lines show cored profiles with $r_c=0.2, 1.0$ and 5 kpc, respectively. The lower panel shows the slope of the mass profiles associated to these models. As expected from eq.~(\ref{eq:gamma}), the values of $\Gamma$ measured in Fornax and Sculptor suggest $r_c\gtrsim 1$ kpc. 

 It is useful to define an ``observationally-relevant'' range of core sizes. 
Motivated by the above measurements, we set a minimum core size comparable to the half-light radius of the dwarf\footnote{Notice also that inferring the presence of tiny ($r_c\ll r_{\rm h}$) DM cores in dSphs is extremely challenging.}, $r_{c,\rm min}\approx r_{\rm h}$. Given that stars are deeply embedded within the CDM halos of dSphs, $r_{\rm h}\sim 0.1 r_s$ (Pe\~narrubia et al. 2008), we consider minimum core sizes $70\le r_{c,\rm {min}}/{\rm pc}\le 370$ for haloes with $M_{\rm vir}/M_\odot\sim 10^{8-11}$. 
The maximum size, although poorly constrained observationally, can be safely assumed to be $r_{c,\rm max}< r_{\rm vir}$. Hence, here we inspect transformations in the range $0.1 \lesssim  r_c/r_s < c$. This uncertainty, together with the details of the outer halo profile (see \S\ref{sec:transf}), has a small relevance to our conclusions.

\subsection{Energetics of the core/cusp transformation}
\begin{figure}
\includegraphics[keepaspectratio=true,width=84mm]{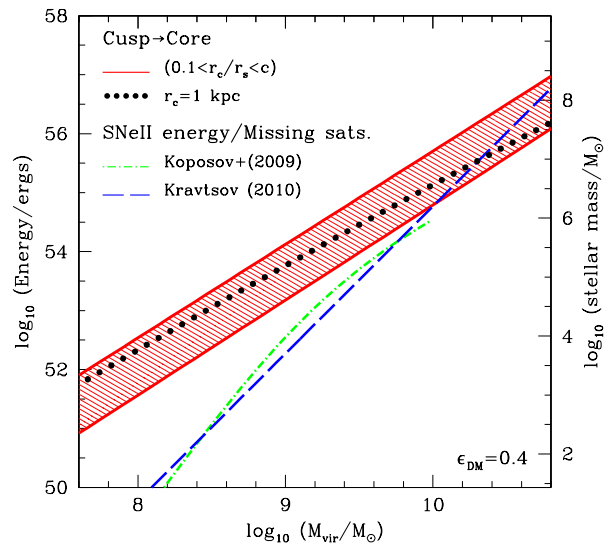}
\caption{Minimum energy ($\Delta E$, red dashed area) required to form a DM core with a size $0.1 \le r_c/r_s \le c$, where $c=r_{\rm vir}/r_s$ is the halo concentration an NFW halo with virial mass $M_{\rm vir}$ at $z=0$. The energy {\it required} to form a DM core with $r_c=1$ kpc is marked with black dots for reference. On the right vertical axis we plot the luminosity derived from a star formation efficiency $F_\star=F_\star(M_{\rm vir})$ tuned to reproduce the number of luminous satellites in our Galaxy. Luminosities are converted into SNeII energy using eq.~(\ref{eq:delwms}) and adopting a strong energy coupling $\epsilon_{\rm DM}=0.4$. Notice that the tension between the `core/cusp' and `missing satellite' problems arise in haloes with $M_{\rm vir}\lesssim 10^{10} M_\odot$.}
\label{fig:work}
\end{figure}

Thick red lines in Fig.~\ref{fig:work} show the range of energies required to remove DM cusps as a function of the halo virial mass. 
As expected from eq.~(\ref{eq:delw}) the amount of energy required by the transformation scales as $\Delta W\propto M_{\rm vir}^2$. At a fixed mass the energy range (shaded area) is relatively narrow, the {\it lower} limit of $\Delta W$ being determined by the minimum core size, $r_{c,\rm {min}}\sim 0.1 r_s$ (see \S\ref{sec:coresize}), and the upper limit being given by the asymptotic limit of $\Theta$ for $x\gg 1$. For ease of reference we also mark the case $r_c=1$ kpc with black dots. 

The CDM haloes of bright dSphs have estimated virial masses in the range $M_{\rm vir}/M_\odot\sim 10^{9-10}$ (Pe\~narrubia et al. 2008). Thus, baryons must produce of the order of $10^{53}\lesssim \Delta W/({\rm erg})\lesssim 10^{55}$ in order to effect the profile transformation on the dSph mass scales.

\section{CDM problems and baryonic feedback}\label{sec:feedback}

\subsection{Supernova energy}\label{sec:sn}
The only plausible source of such an immense amount of energy appears to be the averaged $E_{\rm SN}\sim 10^{51}$ erg of kinetic energy released per SNII explosion (Utrobin \& Chugai 2011). The fraction that contributes to the pressure term in the hydrodynamical equations is the so-called `energy coupling' ($\epsilon_{\rm SN}$). Its value is still a matter or debate. Typical values range from $\epsilon_{\rm SN}=0.01$ (e.g. Kellermann 1989) to 0.40 (Governato et al. 2010). Recently, Revaz \& Jablonka (2012) argue that $\epsilon_{\rm SN}\approx 0.05$ in order to describe the relation between metallicity and luminosity found in dSphs (e.g. Kirby et al. 2011). Note, however, that the fraction of energy transferred into the gas ($\epsilon_{\rm SN}$) may be much larger than the fraction of energy eventually transferred into the DM distribution, which we term $\epsilon_{\rm DM}$. We will adopt $\epsilon_{\rm DM}=0.4$, which gives an upper limit to the energy available, since $\epsilon_{\rm DM}<\epsilon_{\rm SN}$, and a value $\epsilon_{\rm SN}=0.4$ is already quite optimistic (but facilitates comparison with Governato et al results).

The stellar mass ($M_\star$) required to generate an energy $\Delta E$ through SNeII explosions is
\begin{eqnarray}
\label{eq:delwms}
\Delta E= \frac{M_\star}{\langle m_\star\rangle} \xi(m_\star>8 M_\odot)E_{\rm SN}\epsilon_{\rm DM};
\end{eqnarray}
where we assume that stars form following a universal Initial Mass Function (IMF in short), $\xi(m_\star)$; and that only stars with masses $m_\star>8 M_\odot$ undergo SNeII during their last evolutionary stages. For simplicity we adopt a Kroupa (2002) IMF, which gives a fraction of massive single stars $\xi(m_\star>8M_\odot) =0.0037$ and a mean stellar mass $\langle m_\star \rangle=0.4M_\odot$.
 
Eq.~(\ref{eq:delwms}) relates stellar mass to the amount of SNeII energy available to do work on the halo, and we use it to draw the right-vertical axis in Fig.~\ref{fig:work}.  If we allow stellar mass fractions to take values as large as the universal baryon fraction --i.e. $M_\star/M_{\rm vir}=\Omega_b/\Omega_m\simeq 0.047/0.28\simeq 0.16$--then Fig.~\ref{fig:work} indicates that core formation via SNeII feedback is indeed energetically viable on all mass scales. However, this quantity is strongly constrained by the collective properties of the satellite population.

\subsection{The missing satellite problem}\label{sec:missing}
The number of visible structures in our Galaxy constrains the stellar mass that CDM subhaloes can form before being accreted onto the host and becoming ``satellites''. It is useful to define the {\it star formation efficiency} as the fraction of the virial mass that a DM halo converts into stars normalized to the universal baryon fraction, i.e.
\begin{eqnarray}
\label{eq:fs}
F_\star\equiv \bigg(\frac{M_\star}{M_{\rm vir}}\bigg)\bigg(\frac{\Omega_b}{\Omega_m}\bigg)^{-1}.
\end{eqnarray}

Because collision-less CDM simulations do not contain baryons, $F_\star$ is typically ``measured'' by associating the most massive subhaloes to the most luminous satellite galaxies. Due to the divergent nature of the subhalo mass function, $f_s(M_{\rm vir})\propto M_{\rm vir}^{-1.9}$, a strong suppression of star formation in low-mass haloes is needed in order to reconcile the (small) number of visible satellites with the overwhelming number of substructures predicted by CDM (a.k.a the ``missing satellite problem''; Moore et al. 1999; Klypin et al. 1999). E.g.
\begin{eqnarray}
\label{eq:fssat}
F_{\star,{\rm sat}}\approx 10^{-3}\bigg(\frac{M_{\rm vir}}{10^{10}M_\odot}\bigg)^{1/3};
\end{eqnarray}
in haloes with $M_{\rm vir}\lesssim 10^{10}M_\odot$ (Kravtsov 2010; see also Tollerud et al. 2008 and Koposov et al. 2009). Recent studies show that eq.~(\ref{eq:fssat}) is also consistent with the abundance of luminous structures found in external galaxies (e.g. Behroozi et al. 2010).

From eq.~(\ref{eq:delwms}) the formation of a DM core with $r_c\gtrsim 0.1 r_s$ requires 
\begin{eqnarray}
\label{eq:fsmin}
F_{\star,{\rm core}}(z=0)\gtrsim 10^{-3}\bigg(\frac{M_{\rm vir}}{10^{10}M_\odot}\bigg)^{2/3}\bigg(\frac{\epsilon_{\rm DM}}{0.4}\bigg)^{-1}.
\end{eqnarray}
Comparison with eq.~(\ref{eq:fssat}) shows that $F_{\star,{\rm core}}/F_{\star,{\rm sat}}\sim (M_{\rm vir}/10^{10}M_\odot)^{1/3}\lesssim 1$ at $M_{\rm vir}\lesssim 10^{10}M_\odot$, implying that the formation of DM cores in dSphs leads to an over-abundance of luminous satellites (see also Fig.~\ref{fig:work}). 


However, this comparison has an important caveat, viz. that halo masses and concentrations are calculated at $z=0$, whilst abundance-matching techniques return star formation efficiencies at the {\it time of accretion} ($z_{\rm acc}$). For MW-like haloes the accretion redshift of the surviving satellite population peaks at $z_{\rm acc}\approx 1$ (Zentner \& Bullock 2003; Pe\~narrubia \& Benson 2005). 

Let us first inspect the energetics of the core/cusp transformation as a function of redshift. Because the Universe, and therefore the structures that form within, was denser in the past, $\Delta E$ tends to {\it increase} as we go to higher redshifts at a fixed halo mass. This dependence is approximately linear during the matter-dominated era ($z\gtrsim 1$), because the virial over-densities $\delta_{\rm vir}=\rho_{\rm vir}/\rho_m-1\simeq 18 \pi^2$ were essentially independent of redshift, and because the concentration evolves as $c(z)\approx c(z=0) (1+z)^{-1}$ (Wechsler et al. 2002). Hence at a fixed halo mass we find that $F_{\star, {\rm core}}(z)\sim F_{\star, {\rm core}}(z=0)(1+z)$ at $z\gtrsim 1$. 

Unfortunately, the redshift at which the profile transformation takes place (denoted here as $z_{\rm core}$) is unknown. Given that $F_{\star,{\rm sat}}$ is measured at the time of accretion ($z_{\rm acc}$), there remains the possibility that DM cusps were removed at $z_{\rm core}\gg z_{\rm acc}$, so that $M_{\rm vir}(z_{\rm core})\ll M_{\rm vir}(z_{\rm acc})$. Because the energy that baryons must supply scales as $\Delta E\propto M_{\rm vir}^2$, at sufficiently high redshifts the formation of cored DM profiles becomes less energetically demanding, even though DM haloes were on average denser. Below we inspect this possibility in detail.

\begin{figure}
\includegraphics[keepaspectratio=true,height=84mm]{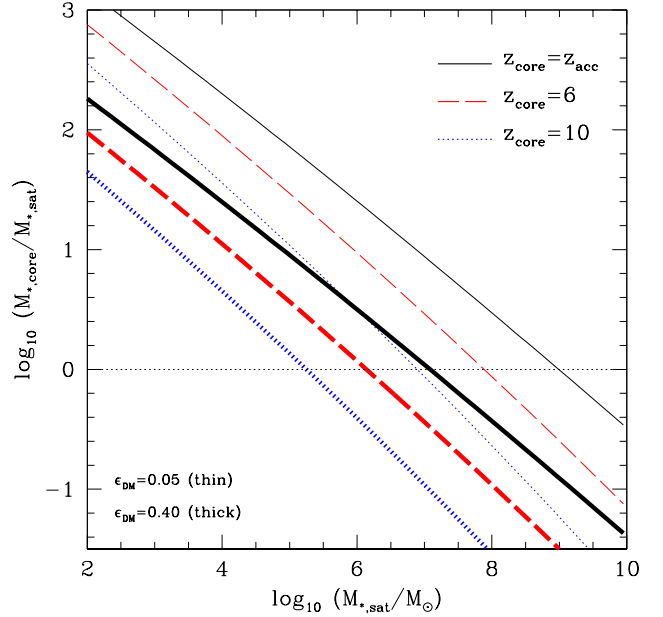}
\caption{Minimum stellar mass required to form a DM core with $r_c=1$ kpc, $M_{\star,{\rm core}}$, against the stellar mass tuned to reproduce the number of visible substructures in CDM simulations, $M_{\star,{\rm sat}}$. Solid lines show both quantities calculated at an accretion time $z_{\rm acc}= 1$, which roughly corresponds to the time at which the majority of the surviving satellites are accreted onto MW-like haloes. Dotted and short-dashed lines show how the stellar masses compare if the cusp removal is shifted to earlier times in the mass evolution of the satellites ($z_{\rm core}=6$ and 10, respectively). Thin and thick lines adopt $\epsilon_{\rm DM}=0.05$ and 0.40, respectively.}
\label{fig:sf}
\end{figure}

\subsection{Transformation redshift}\label{sec:sims}
Fig.~\ref{fig:sf} compares the minimum stellar mass required to form a DM core with $r_c=1$ kpc, $M_{\star,{\rm core}}$, against the stellar mass tuned to reproduce the number of visible substructures in CDM simulations, $M_{\star,{\rm sat}}$ (eq.~\ref{eq:fssat}). 

Models that reconcile the `missing satellite' {\it and} `core/cusp' problems by suppression star formation obey the condition $M_{\star,{\rm core}}\lesssim M_{\star,{\rm sat}}$ and fall below the horizontal dotted line in Fig.~\ref{fig:sf}. Notice that if DM cusps are removed by the time satellites are accreted ($z_{\rm core}=z_{\rm acc}\approx 1$; solid line) the presence of DM cores should be limited to galaxies with $M_\star\gtrsim 10^7 M_\odot$ for $\epsilon_{\rm DM}=0.40$ (thick lines), in agreement with the hydro-dynamical simulations of G12.

The threshold between cored and cuspy DM profiles moves to fainter luminosities if the transformation is shifted to earlier times (see \S\ref{sec:missing}). How early in the past can it be shifted? According to Pontzen \& Governato (2012) and Teyssier et al. (2012) cusps are removed on a time-scale of $\sim 1$ Gyr, which roughly corresponds to the time between multiple in- and out-flows of gas during subsequent SNeII explosions. Thus in a $\Lambda$CDM cosmogony $z_{\rm core}\sim 6$ appears the earliest plausible time for the transformation, whereas $z\sim 10$ corresponds to the earliest epoch of star formation in low-mass haloes (Bovy \& Dvorkin 2012). 

Dotted and short-dashed lines in Fig.~\ref{fig:sf} show the effects of shifting the cusp removal at $z_{\rm core}=6$ and 10, respectively. In order to derive conservative estimates we assume no further star formation after the DM core is formed. Halo masses are evolved back in time using Wechsler et al. (2002) formalism. These authors show that (i) CDM haloes grow exponentially, so that $M_{\rm vir}(a_{\rm core})\approx M_{\rm vir}(a_{\rm acc})\exp[-2 a_c(a_{\rm acc}/a_{\rm core}-1)]$, where $a=(1+z)^{-1}$ is the scale factor; and (ii) the ``collapse-time'' of the halo ($a_c$) correlates tightly with the virial concentration, so that $a_c\sim 4.1 a_{\rm acc}/c(z_{\rm acc})$ independent of the redshift when the halo is observed.

Fig.~\ref{fig:sf} shows that although shifting the transformation to $z_{\rm core}\approx 6$ helps to accommodate DM cores in the bright dSphs (i.e. $M_\star\gtrsim 10^6 M_\odot$), the tension cannot be completely eliminated but is simply shifted to lower luminosities. Notice also that the formation of cored halo profiles in dSphs ($M_{\star}\lesssim 10^{7}M_\odot$) requires $\epsilon_{\rm DM}\approx \epsilon_{\rm SN}\sim 1$.

\section{Discussion}\label{sec:diss}
Our estimates highlight a tension between CDM predictions and observations on small galactic scales. 
On the one hand, we find that the cored density profile measured in two of the brightest MW dSphs points toward an efficient conversion of primordial gas into stars on the mass scales of dSphs ($M_{\rm vir}\lesssim 10^{10}M_\odot$). On the other, a strong suppression of star formation is required {\it on the same mass scales} in order to accommodate the small number of visible satellites with the halo mass function predicted by collisionless CDM simulations.

It is useful to identify conditions whereby this tension may be eased and discuss their compatibility with observations. (i) The removal of DM cusps requires a very efficient transfer of supernova energy into DM particles (i.e. $\epsilon_{\rm DM}\approx \epsilon_{\rm SN}$). The results of Pontzen \& Governato (2012) suggest that the required efficiency may be reached if starbursts are sufficiently frequent and energetic. (ii) The supernova energy coupling ($\epsilon_{\rm SN}$) must be order unity. However, such a strong coupling may be incompatible with the observed luminosity-metallicty relationship of dSphs (Revaz \& Jablonka 2012) (iii) Cusps must be removed at high redshifts. The cored profiles found in Sculptor and Fornax suggest $z_{\rm core}\gtrsim 6$. However, stellar ages are at odds with this scenario, as both galaxies show extended periods of star formation continuing for 6-7 Gyr in Sculptor (de Boer et al. 2012a), and for 12-13 Gyr in Fornax (del Pino et al. 2011; de Boer et al. 2012b). Also, even if DM cores formed when their haloes had tiny masses, subsequent hierarchical accretion of unperturbed subhaloes would tend to re-grow DM cusps (Dehnen 2005).
(iv) A possibility not explored here is the case of a top-heavy IMF. A higher fraction of massive stars would increase the feedback energy reservoir without aggravating the over-abundance of luminous substructures. 

 
An alternative route out of the tension is for the satellite luminosity function to be substantially reshaped by mechanisms other than suppression of galaxy formation. Recent hydro-dynamical simulations show that baryonic feedback lowers the number of surviving satellites with respect to the DM-only case (Zolotov et al. 2012). This difference is partially due to the enhanced mass-loss rate of satellites embedded in cored DM haloes (Pe\~narrubia et al. 2010). Baryonic feedback may therefore ease the constraints discussed above by increasing the tidal disruption rate of massive (bright) subhaloes moving on eccentric orbits. 
The fact that the star formation efficiencies of the bright MW satellites lie above those dictated by abundance matching arguments (Sawala et al. 2012; Boylan-Kolchin et al. 2012) may point in this direction.  

Our results also provide compelling reasons to push measurements of halo mass profiles to galaxies with $M_*\lesssim 10^6M_{\odot}$, as the tension between `core/cusp' and `missing satellites' problems only increases toward lower luminosity.

\section*{Acknowledgments}
JP acknowledges support from the Ram\'on y Cajal Program as well as by the Spanish grant AYA2010-17631. AP is supported by the Oxford Martin School. MGW is supported by NASA through Hubble Fellowship grant HST-HF-51283.01-A.

{}

\end{document}